\documentclass{ws-mpla}
\usepackage{amsmath}
\usepackage{graphicx}
\usepackage{amsfonts}
\usepackage{amssymb}

\begin{document}

\markboth{Simone Mercuri and Giovanni Montani}
{Dualism between Physical Frames and Time in Quantum Gravity}

%
\catchline{}{}{}{}{}
%

\title{DUALISM BETWEEN PHYSICAL FRAMES AND TIME IN QUANTUM GRAVITY}

\author{\footnotesize SIMONE MERCURI}

\address{ICRA International Center for Relativistic Astrophysics c/o Dipartimento di Fisica (G9) Universit\`a di Roma ``La Sapienza'', Piazza A.Moro 5 00185 Rome, Italy.\\
mercuri@icra.it}

\author{GIOVANNI MONTANI}

\address{ICRA---International Center for Relativistic Astrophysics c/o Dipartimento di Fisica (G9) Universit\`a di Roma ``La Sapienza'', Piazza A.Moro 5 00185 Rome, Italy. \\Dipartimento di Fisica Universit\`a di Roma ``La Sapienza''. \\montani@icra.it}

\maketitle

\pub{Received (Day Month Year)}{Revised (Day Month Year)}

\begin{abstract}
In this work we present a discussion of the existing links between the
procedures of endowing the quantum gravity with a real time and of including
in the theory a physical reference frame.\newline More precisely, as first
step, we develop the canonical quantum dynamics, starting from the Einstein
equations in presence of a dust fluid and arrive to a Schr\"odinger evolution.
Then, by fixing the lapse function in the path-integral of gravity, we get a
Schr\"odinger quantum dynamics, of which eigenvalues problem provides the
appearance of a dust fluid in the classical limit.

The main issue of our analysis is to claim that a theory, in which the time
displacement invariance, on a quantum level, is broken, is indistinguishable
from a theory for which this symmetry holds, but a real reference fluid is included.

\keywords{Quantum gravity; reference frame.}
\end{abstract}

\ccode{PACS Nos.: 04.60.-m}

\section{General Statements}	

Gravitational interaction in view of its non linear nature has to find its
most appropriate quantum approach in the path-integral representation, in fact
there are many formal analogies between gravitational and non Abelian gauge theories, especially when the former is represented in a connection formalism
\cite{Ash1986}. Furthermore in General Relativity there is no chance to speak
of a well defined perturbation theory, since free asymptotic states in strict
sense do not exist, in fact unlike Yang Mills fields, the gravitational one is
the geometry of the space-time and should couple locally with any physical
system. So that the notion of ``in'' and ``out'' states loses its physical
meaning. However it is worth noting that a satisfactory description of the
quantum field theory on a fixed background including the graviton-matter
interaction is provided by the modern approach of string theory \cite{Pol1998,GSW1987}; nevertheless to extend the quantization procedure to the whole
geometry of the space-time is a problem which, up to now, did not find a
solution within the superstring framework.\newline Promising formalisms toward
the non perturbative quantization of the space-time structure appeared in the
first nineteenth, when the \emph{loop quantum gravity} theory, \cite{ARS1991,Rov1997,Thi2001}, was settled down. Indeed the spin networks formalism
provides well defined states for the canonical quantum gravity theory and can
be upgraded to a path-integral procedure by virtue of the recent \emph{spin
foam} approach \cite{BC1997,Ori2001}.

The first attempt for a path-integral approach to quantum gravity was due to
Hawking et al. \cite{GH1977,Haw1978,HH1983} in the early eighties. They
addressed this problem from a rather heuristic point of view, in fact their
path-integral had not a definite Lebesgue measure and even its integrand was
ill defined.

Our point of view is that one of the main problems in both the canonical and path-integral approaches to quantum gravity is the non existence of evolution for the quantum dynamics. We think that this problem is strictly connected with the choice of a physical reference system. \newline
The aim of this work is, in fact, to show how there exists a dualism, in quantum
gravity, between introducing a real reference fluid and breaking down the
invariance of the theory under time displacements. The contact issue in these
two points of view consists in providing a physical time for the quantum
dynamics, \cite{KuT1991,Rov1991,Ish1992,BrKu1994,BiKu1997,Mon2002,MM2003,Nik2003}, (for applications see \cite{VW1998,VWS2000,Mon2003}%
).\newline To achieve this result, we first show, in ``section 2'', that to
quantize the gravitational field in presence of a dust reference fluid implies
to define an appropriate time variable as the conjugate momentum to the
eigenvalue of the super-Hamiltonian. Indeed such a non-zero eigenvalue is due
to the dust contribution and, being in general non positive, allows us to
claim that: \emph{a reference fluid, in quantum gravity, is never a test one
and its energy density can be even negative}.\newline Then, in ``section 3'',
the picture is completed by fixing the lapse function in the path-integral for
the gravitational field, so outlining that it leads, in first approximation,
to a Schr\"{o}dinger dynamics. Via the eigenvalues problem, we finally arrive,
in the classical limit, to show the appearance of a dust fluid.

In section 4 brief concluding remarks follow.

\section{Gravity-Fluid Quantization}

In treating General Relativity, we commonly refer to a reference frame as to a
specific system of coordinates. However, these two notions are significantly
different, because it is necessary, to speak of a real reference frame, the
introduction in the space of a material framework (at least locally), to which
physically link the description of the phenomena. In General Relativity a real
reference frame is implemented considering a fluid which fills the whole
space-time (or isolated regions) by its world lines; of course more general
physical entities can be used to define a reference frame, but they can be
reduced to the primitive notion of a fluid.\newline To specify a reference
fluid we have to assign a 4-velocity vector field describing the world lines
of each fluid element; being a physical system, the fluid has always
associated a non zero energy-momentum tensor. Indeed, it is just in neglecting
the energy-momentum carried by the fluid the main source of confusion between
the notions of reference frame and of coordinates system. The latter can be
regarded as simply labeling the points of $\mathcal{M}^{4}$ and it is well
implemented by a ``test fluid''. Let us now emphasize a crucial difference
existing between the classical and quantum dynamical role played by a real
reference fluid. Below we will discuss the case of a \emph{dust}, which plays
the role of reference fluid. In view of its phenomenological nature, the
canonical quantization procedure is developed by inferring the Hamiltonian
constraints from the Einstein equations. \newline Let us consider a
4-dimensional space-time manifold, $\mathcal{M}^{4}$, on which a coordinates
system $\{y^{\mu}\}$ and a metric tensor $g_{\mu\nu}(y^{\rho})$ (with the
signature $-+++$, where $\mu,\nu,\rho=0,1,2,3$) are assigned. A dust fluid is
characterized by its proper energy density $\varepsilon(y^{\rho})$ and the
4-velocity $u^{\mu}(y^{\rho})$ ($g_{\mu\nu}u^{\mu}u^{\nu}=-1$), which leads to
the energy-momentum tensor $T_{\mu\nu}=\varepsilon u_{\mu}u_{\nu}$%
.\newline The Einstein equations and the conservation laws, for the coupled
gravity-fluid system take the form:
\begin{align}
&  G_{\mu\nu}=\chi\varepsilon u_{\mu}u_{\nu},\label{xy1}\\
&  u^{\nu}\nabla_{\nu}u^{\mu}=0,\label{2}\\
&  \nabla_{\nu}(\varepsilon u^{\nu})=0,
\end{align}
where $G_{\mu\nu}$ and $\chi$ denote respectively the Einstein tensor and
constant.\newline We can decompose the metric tensor as follows:
\begin{equation}
g_{\mu\nu}=h_{\mu\nu}-u_{\mu}u_{\nu}\Rightarrow h_{\mu\nu}u^{\nu}=0.
\label{xy2}%
\end{equation}
Remembering a well-known result, it is easy to show that the following
relations take place \cite{Wal1984,Thi2001}
\begin{align}
G_{\mu\nu}u^{\mu}u^{\nu}  &  =-\frac{H(h_{ij},p^{ij})}{2\sqrt{h}}%
=\chi\varepsilon,\label{xy3}\\
G_{\mu\nu}u^{\mu}h_{i}^{\nu}  &  =\frac{H_{i}(h_{ij},p^{ij})}{2\sqrt{h}}=0\,.
\end{align}
Here $h_{ij}$ ($ij=1,2,3$) denotes the 3-metric of the spatial hypersurfaces
orthogonal to $u^{\mu}$ (requiring the condition $u_{[\mu}\nabla_{\nu}%
u_{\rho]}=0$ the Frobenius theorem (see \cite{Wal1984} p.434) assures the existence
of the hypersurfaces orthogonal to $u_{\mu}$), $h\equiv deth_{ij}$ and
$p^{ij}$ its conjugate momentum, while $H$ and $H_{i}$ refer respectively to
the super-Hamiltonian and the super-momentum for the gravitational field.
Indeed, because of its phenomenological nature we do not deal with the
Lagrangean formulation for the fluid and, therefore, the above relations hold
if we make the reasonable assumption that the conjugate momentum $p^{ij} $ be
not affected by the matter variables (i.e. the fluid term in ADM formalism
should not contain the time derivative of the 3-metric tensor). These
equations remain valid in any system of coordinates, being the dependence on
the lapse function and the shift vector contained in the remaining Einstein
equations. However only the Hamiltonian constraints are relevant for the
quantization procedure and, in the comoving frame, when the 4-velocity becomes
$u^{\mu}=\{1,\mathbf{0}\}$ ($N=1\;N^{i}=0$), we have to retain also the
conservation law
\begin{equation}
\varepsilon\sqrt{h}=-\dfrac{\omega(x^{i})}{2\chi}, \label{xy5}%
\end{equation}
where $x^{i}$ denotes the spatial coordinates of the comoving frame and
$\omega(x^{i})$ a generic 3-scalar density of weight 1/2. Here, a crucial
point relies on the synchronous nature of the comoving frame as a consequence
of the geodesic motion of the dust fluid ((\ref{2}) reduces to an identity,
because of the comoving form of $u^{\mu}$). \newline Thus, when the
coordinates system becomes a real physical frame, the Hamiltonian constraints
read
\begin{equation}
H=\omega(x^{i}),\quad H_{i}=0. \label{xy6}%
\end{equation}
Now, to assign a Cauchy problem for such a system, in which equations
(\ref{xy6}) play the role of constraints on the initial data, corresponds to
provide on a (non-singular) space-like hypersurface, say $\Sigma^{(0)}$, the
values $\{h_{ij}^{(0)},p^{(0)ij},\varepsilon^{(0)}\}$; $\omega^{(0)}$ can be
calculated, by (\ref{xy6}), from these values .\newline It follows that, by
specifying a suitable initial condition, the value of $\omega^{(0)}$ can be
made arbitrarily small; from the constraints point of view, a very small value
of $\omega^{(0)}$ means, where $h^{(0)}$ is not so, that the fluid becomes a
test one (being $\omega$ a constant of the motion). We emphasize that for
finite values of $\omega$, $h$ should not vanish to avoid unphysical diverging
energy density of the fluid.

The canonical quantization of this system is achieved as soon as we implement
the canonical variables into quantum operators and annihilate the state
functional $\Psi$ via the Hamiltonian operator constraints. Thus the quantum
dynamics obeys the following eigenvalue problem:
\begin{equation}
\widehat{H}\Psi(\{h_{ij}\},\omega)=\omega\Psi(\{h_{ij}\},\omega), \label{xy7}%
\end{equation}
\newline where $\{h_{ij}\}$ refers to a whole class of 3-geometries, so that
the super-momentum constraint automatically holds.\newline In the above
equation (\ref{xy7}), the spatial function $\omega$ plays the role of the
super-Hamiltonian eigenvalue; in this respect, we observe how its values can
no longer be assigned by the initial conditions, but they have to be
determined via the spectrum of $\widehat{H}$. We conclude that, in the quantum
regime, a real dust reference fluid never approaches a test system and, in
view of the super-Hamiltonian structure (the supermetric has no definite
sign), its energy density is not always positive.\newline In order to
understand the physical meaning of including a real dust fluid in quantum
gravity, let us take the Fourier transform of $\Psi$ as
\begin{equation}
\chi(\{h_{ij}\},\tau)=\int D\omega\Psi(\{h_{ij}\},\omega)exp\{-\frac{i}{\hbar
}(\omega\tau)\}\,, \label{xy8}%
\end{equation}
denoting by $D\omega$ the measure and by $\tau$ a continous space
function.\newline In view of equation (\ref{xy8}), the functional $\chi$ must
satisfy the Schr\"{o}dinger-like equation
\begin{equation}
i\hbar\partial_{\tau}\chi(\{h_{ij}\},\tau)=\widehat{H}\chi(\{h_{ij}\},\tau).
\label{xy9}%
\end{equation}
The Schr\"{o}dinger-like character of the above equation, allows us to infer
that the quantum dynamics of the gravitational field acquires a time
evolution, as soon as, it is constrained to physical frames.

\section{Path-Integral Approach to Hyper Time}

In this section we introduce a real time in the quantum theory by defining a
path-integral in which the lapse function is fixed; then we show how the
eigenvalues problem associated with the Schr\"{o}dinger evolution provides, in
the semi classical limit, the energy density of a dust reference fluid. To
this end, we adopt for the gravitational field the following path-integral:
\begin{equation}
\left|  h_{ij}^{\left(  2\right)  },\Sigma_{\left(  2\right)  }^{3}%
\right\rangle =\int Dh^{\left(  1\right)  }Dp^{\left(  1\right)  }%
\int\limits_{\mathcal{F}}D\Omega\exp\dfrac{i}{\hbar}\left\{  \left(
S+S^{\star}\right)  +2\textstyle\int\limits_{\partial M^{4}}d^{3}x\sqrt
{h}k\right\}  \left|  h_{ij}^{\left(  1\right)  },\Sigma_{\left(  1\right)
}^{3}\right\rangle , \label{pi}%
\end{equation}
calculated on all the possible path which connect the 3-metric $h_{ij}%
^{\left(  1\right)  }$ on the hypersurface $\Sigma_{\left(  1\right)  }^{3}$
to the 3-metric $h_{ij}^{\left(  2\right)  }$ on the hypersurface
$\Sigma_{\left(  2\right)  }^{3}$, moreover an integration over all the possible
initial configuration for the 3-metric is performed. In (\ref{pi}) $S$ is the
ADM action of the gravitational field and we have added the new term:
\begin{equation}
S^{\star}=\dfrac{1}{16\pi}\int\limits_{t_{0}}^{t}dt^{^{\prime}}\int
\limits_{\Sigma_{t^{^{\prime}}}^{3}}d^{3}x\sqrt{h}\Lambda\left(  N-N^{\ast
}\right)  ; \label{sstar}%
\end{equation}
which allows us to fix the Lapse function to the fixed value $N^{\ast},$
through a Lagrange multipliers $\Lambda;$ in view of its structure such a term
satisfies the appropriate constraints algebra:
\begin{equation}
\left\{  H_{i}\left(  x\right)  ,H_{k}\left(  y\right)  \right\}
=H_{k}(x)\delta_{,i}\left(  x-y\right)  -\left(  ix\rightarrow ky\right)  ,
\end{equation}%
\begin{equation}
\left\{  H_{k}\left(  x\right)  ,\tilde{H}\left(  y\right)  \right\}
=\tilde{H}\left(  x\right)  \delta_{,k}\left(  x-y\right)  ,
\end{equation}%
\begin{equation}
\left\{  \tilde{H}\left(  x\right)  ,\tilde{H}\left(  y\right)  \right\}
=H^{k}\left(  x\right)  \delta_{,k}\left(  x-y\right)  -\left(  x\rightarrow
y\right)  ,
\end{equation}
\newline where $\tilde{H}=H-\sqrt{h}\Lambda.$ Finally we include a boundary
term, $S_{b},$ because we assume non zero normal derivatives on the compact
the 3-hypersurfaces $\Sigma_{t}^{3}.$ \newline Our path-integral describes a
quantum theory in which the time displacements invariance is explicitly
broken, the reason for this approach relies on its equivalence with including
a reference fluid in the dynamics. The meaning of the coordinates $\left(
t,x^{i}\right)  $ will turn out at the end of our procedure, being related to
the fluid labels.

The functional integral is taken over the domain $\mathcal{F}$, which is the
product of the functional spaces corresponding to the 3-metric with non
vanishing determinant ($h\neq0),$ to the conjugate momentum, lapse function,
shift vector and $\Lambda$'s; therefore for the Lebesgue measure we have
$D\Omega\equiv D\Lambda Dp^{kl}Dh_{ij}DN^{a}DN$ (in what follows, if not
complete, the Lebesgue measure will appear with lower labels of the variables
to which it refers).

As soon as we evaluate the functional integral over $\Lambda$, we arrive to
the new path-integral:
\begin{align}
&  \left|  h_{ij}^{(2)},\Sigma_{(2)}^{3}\right\rangle =\int Dh^{\left(
1\right)  }Dp^{\left(  1\right)  }\int\limits_{\mathcal{F}}D\Omega\exp
\dfrac{i}{\hbar}\left\{  S+S^{\star}+S_{b}\right\}  \left|  h_{ij}%
^{(1)},\Sigma_{(1)}^{3}\right\rangle =\nonumber\\
&  =\int Dh^{\left(  1\right)  }Dp^{\left(  1\right)  }\int
\limits_{\mathcal{F}_{phNN^{i}}}D\Omega_{phNN^{i}}\delta\left[  \sqrt
{h}\left(  N-N^{\ast}\right)  \right]  \exp\dfrac{i}{\hbar}\left\{
S+S_{b}\right\}  \left|  h_{ij}^{(1)},\Sigma_{(1)}^{3}\right\rangle
=\nonumber\\
&  =\int Dh^{\left(  1\right)  }Dp^{\left(  1\right)  }\int
\limits_{\mathcal{F}_{phN^{i}}}D\Omega_{phN^{i}}\left(  \prod_{t,x^{k}}%
\dfrac{1}{\sqrt{h}}\right)  \exp\dfrac{i}{\hbar}\left\{  S(N^{\ast}%
)+S_{b}\right\}  \left|  h_{ij}^{(1)},\Sigma_{(1)}^{3}\right\rangle ,
\label{npi}%
\end{align}
where $\prod\limits_{t,x^{i}}\left(  h\right)  ^{-1/2},$ in the last line,
comes out from the evaluation of the delta function and $S_{b}$ denotes the
boundary term.

Since now the wave functional will depend on $N^{\ast}$ explicitly, then it
changes passing from a 3-hypersurface to a neighboring one, so acquiring a
``time'' evolution. We will indicate the parameter of this evolution with the
label ``$t$''. Therefore taking the two hypersurfaces corresponding to $t$ and
$t+\varepsilon$ ($\varepsilon\ll1$), which are two different instants of
``time'' referred to two neighboring hypersurfaces, i.e. to two values of the
function $N^{\ast}.$ The expansion of the left and right side of the equation
(\ref{npi}) leads to:
\begin{align}
&  \left|  h_{ij}^{t+\varepsilon},\Sigma_{t+\varepsilon}^{3}\right\rangle =\left|
h_{ij}^{t},\Sigma_{t}^{3}\right\rangle +\varepsilon\partial_{t}\left|
h_{ij}^{t},\Sigma_{t}^{3}\right\rangle =\nonumber\\
& \nonumber\\
&  =\dfrac{1}{A}\int Dh^{\left(  1\right)  }Dp^{\left(  1\right)  }%
\int\limits_{\mathcal{F}_{p}}D\Omega_{pN^{i}}\left(  1-\frac{i}{\hbar
}\varepsilon\mathcal{H}\right)  \exp\left\{  \dfrac{i}{\hbar}\int d^{3}%
xp^{ij}\left(  h_{ij}^{t+\varepsilon}-h_{ij}^{t}\right)  \right\}  \left|
h_{ij}^{t},\Sigma_{t}^{3}\right\rangle ,
\end{align}
where: $\dfrac{1}{A}=\int\limits_{\mathcal{F}_{h}}Dh\prod\limits_{t,x^{i}%
}\left(  h\right)  ^{-1/2}\exp\left\{  -{i}S_{b}/{\hbar}\right\}$ (for neighboring hypersurfaces $\Sigma_{t}^{3}$ and
$\Sigma_{t+\varepsilon}^{3}$ the functional integral over $\mathcal{F}_{h}$
reduces to a fixed value) and we
approximate the time derivative of the variable $h_{ij}$ as $\partial
_{t}h_{ij}\longrightarrow({h_{ij}^{t+\varepsilon}-h_{ij}^{t}})/{\varepsilon}.$
Now, replacing the momentum terms by the corresponding functional derivatives,
i.e.
\begin{align}
-i\hbar\dfrac{\delta}{\delta h_{ij}^{t}}&\int Dp\exp\left\{  \dfrac{i}{\hbar
}\int d^{3}xp^{ij}\left(  h_{ij}^{t+\varepsilon}-h_{ij}^{t}\right)  \right\}=
\nonumber\\&=\int Dpp^{ij}\exp\left\{  \dfrac{i}{\hbar}\int d^{3}xp^{ij}\left(
h_{ij}^{t+\varepsilon}-h_{ij}^{t}\right)  \right\}  ,
\end{align}
the leading term of the expansion above leads to the Schr\"{o}dinger-like
equation:
\begin{equation}
i\hbar\partial_{t}\left|  h_{ij}^{t},\Sigma_{t}^{3}\right\rangle
=\widehat{\mathcal{H}}\left|  h_{ij}^{t},\Sigma_{t}^{3}\right\rangle ,
\label{eqs}%
\end{equation}
where $\widehat{\mathcal{H}}=\textstyle\int\limits_{\Sigma_{t}^{3}}%
d^{3}x\left(  N^{\ast}\widehat{H}+N^{k}\widehat{H}_{k}\right)  $. In the above
formula (\ref{eqs}) as well as in (\ref{xy9}) the Hermitian character of
$\widehat{H}$ requires the same normal ordering introduced in \cite{Mon2002,MM2003}. Furthermore in
analogy to the proof presented in \cite{HH1983}, we can get $\widehat{H}%
_{k}\left|  h_{ij}%
^{t},\Sigma_{t}^{3}\right\rangle=0$. Now defining the wave functional $\Psi\left(  t,\left\{
h_{ij}\right\}  \right)$ (where with the notation $\left\{
h_{ij}\right\}$ we indicate the 3-geometries) as the projection of the ket $\left|  h_{ij}%
^{t},\Sigma_{t}^{3}\right\rangle $ on the functional space, it is easy
to realize that if we expand $\Psi\left(  t,\left\{
h_{ij}\right\}  \right)$ as follows:
\begin{equation}
\Psi\left(  t,\left\{  h_{ij}\right\}  \right)  =\underset{y_{t}^{\ast}}{\int
}D\omega\Theta\left(  \omega\right)  \xi_{\omega}(\{h_{ij}\})\exp\left(
-\dfrac{i}{\hbar}\int\limits_{t_{0}}^{t}dt^{\prime}\underset{\Sigma_{t}^{3}%
}{\int}d^{3}x\left(  N\omega\right)  \right)  , \label{expan}%
\end{equation}
then the following eigenvalues problem takes place:
\begin{equation}
\widehat{H}\xi_{\omega}(\{h_{ij}\})=\omega\left(  x^{j}\right)  \xi_{\omega
}(\{h_{ij}\}).
\end{equation}
Above we regarded $\xi_{\omega}$ as taken on the 3-geometries in order to
consider the constraint $\widehat{H}_{k}\xi_{\omega}=0$. Now, toward the
classical limit, we choose the wave functional in the form: $\psi\propto
\exp\left\{  i\sigma\right\}  /\hbar$; then interpreting the ${\delta\sigma
}/{\delta h_{ij}}$ as the conjugate momentum $p^{ij}$ (this is allowed by the
Hamilton-Jacobi structure of the leading order term in $\hbar$), we easily
recognized that the above eigenvalues problem reduces to the classical
Hamiltonian constraints:
\begin{equation}
H=\omega,\qquad H_{k}=0.
\end{equation}
Since the following relations hold
\begin{equation}
G_{\mu\nu}u^{\mu}u^{\nu}=-\dfrac{H}{2\sqrt{h}}=-\dfrac{\omega}{2\sqrt{h}},
\end{equation}%
\begin{equation}
G_{\mu\nu}u^{\mu}h_{i}^{\nu}=\frac{H_{i}}{2\sqrt{h}}=0,
\end{equation}
(where $u^{\mu}$ is the orthonormal vector to the 3-hypersurfaces), then in
the comoving reference ($u^{\mu}=(1,\mathbf{0)}$, i.e. $N=1,N^{i}=0$) we may
write the $0-0$ and $0-j$ Einstein equation:
\begin{equation}
G_{00}=\chi\varepsilon,\quad G_{0j}=0,
\end{equation}
with $\varepsilon=-{\omega}/{2\chi\sqrt{h}}$ and by (\ref{xy2}) $h_{i}^{\mu
}=\delta_{i}^{\mu}$. Finally the variation of $S^{\ast}$ with respect to
$h_{ij}$ does not contribute any term to the Einstein equation, because the
variation with respect to $\Lambda$ requires $N=N^{\ast}$; therefore we may
also write $G_{ij}=0$. By the covariance principle of Einstein theory, it
follows that in a generic coordinates system, we should re obtain the
equations (\ref{xy1}). We adopted exactly the same letters ($u^{\mu
},\varepsilon,\omega$) to denote the quantity of previous and present section,
just to stress the complete equivalence of the two opposite paths: from fluid
to Schr\"{o}dinger equation, from Schr\"{o}dinger equation to fluid. We
conclude this section by stessing that the meaning of the coordinates $\left(
t,x^{i}\right)  $ in the Schr\"{o}dinger equation (\ref{eqs}) is provided by
the fluid interpretation. In fact for $N=1,$ $N^{i}=0,$ such coordinates
coincide respectively with the ``clock'' and the elements of the fluid, while
in general they are reparameterization of such physical coordionates.

\section{Concluding Remarks}

The most puzzling feature of the Wheeler-DeWitt equation consists in its
\emph{frozen formalism}, i.e. of the absence of any evolution along the family
of hypersurfaces $\Sigma_{t}^{3}$, which fulfills the whole space-time. Such a
feature is a direct consequence of implementing the time displacements
invariance of the theory and prevents a consistent interpretation of the
resulting quantum dynamics. Here we have shown how if we break down this
invariance on a quantum level, then, in the classical limit, we approach an
Einstein dynamics in presence of a dust reference fluid. Our proof is based on
the equivalence between a Schr\"{o}dinger-like dynamics and the Einstein dust
equations; this feature suggests that, in quantum gravity, the phenomenology
has to be linked to a physical frame. To say better, in quantum regime makes
no longer sense to speak simply of coordinates system, because any fluid can
not be a test one. \newline A subtle consequence of such a statement is the
appearance of a non positive energy density in the theory. Though it calls
attention for deeper understanding, we claim that this feature is a natural
implication of the super-Hamiltonian structure and should not be rejected as
non physical one.\newline Finally, we observe that the equivalence of the
approaches presented in sections 2 and 3 could not seem complete, because in
the former we have in (\ref{xy9}) a local Schr\"{o}dinger dynamics, involving
a physical time $\tau$, while in the latter the dynamics is smeared and the
label time $t$ appears (\ref{eqs}). But, by smearing equation (\ref{xy9})
times $N^{\ast}$ and using the super-momentum constraint, we find:
\begin{equation}
i\hbar\int\limits_{\Sigma_{t}^{3}}d^{3}x\frac{\delta\chi}{\delta\tau}N^{\ast
}=\widehat{\chi}.
\end{equation}
Hence observing that, in view of the construction of section 2, $\tau$ denotes
a synchronous time, it follows the relation $N^{\ast}=\partial_{t}\tau$.
Therefore, the complete equivalence of the two schemes, relies on the natural
identification $\partial_{t}(...)\equiv\int\limits_{\Sigma_{t}^{3}}d^{3}%
x\frac{\delta(...)}{\delta\tau}\partial_{t}\tau.$ We conclude by stressing how
the smeared dynamics got in this work, exactly coincides with that one
presented in \cite{Mon2002}, which was based on the presence of the
Kinematical term \cite{Kuc1981,Mon2002,MM2003} in the gravity
action.\newline The convergence of these two different approaches support the
validity of their physical implications. A final point calls attention for a
deeper understanding: The theory developed in section 3 spontaneously breaks
the time displacements invariance, while to include the refence fluid as in
section 2 preserves the 4-diffeomorphisms; how can these two approaches be equivalent?

The answer to this question relies on the statement of section 2, that a
reference fluid in quantum gravity can never be a test one; thus, on a
quantumn level, only systems of coordinates associated with fluids, which are
source for the space-time curvature, are admissible. If we deal with a single
fluid coupled to the 3-metric then it becomes a ``privileged'' one, because
all other systems of coordinates are simple reparameterization of the physical
coordinates linked to this fluid. On a classical level the covariance is
restated because any other system of coordinates is a physical frame, when
thought as a test fluid. In this sense to consider a reference fluid in
quantum gravity is equivalent to break down the time diffeomorphisms and
provide a time evolution for the dynamics.

\end{document}